\documentclass{article}

\usepackage{PRIMEarxiv}

\usepackage[utf8]{inputenc} 
\usepackage[T1]{fontenc}    
\usepackage{url}            
\usepackage{booktabs}       
\usepackage{amsfonts}       
\usepackage{nicefrac}       
\usepackage{microtype}      
\usepackage{lipsum}
\usepackage{multirow}
\usepackage{fancyhdr}       
\usepackage{graphicx}       
\graphicspath{{media/}}     
\usepackage[style=apa, backend=biber]{biblatex}  
\DeclareLanguageMapping{english}{english-apa}
\usepackage{graphicx}       
\usepackage{url}

\title{Evaluating approaches to identifying research supporting the United Nations Sustainable Development Goals}

\author{Yury Kashnitsky    \\
  Elsevier, the Netherlands                          \\
  \texttt{y.kashnitskiy@elsevier.com}                \\\And
  Guillaume Roberge             \\
  Elsevier, Canada                                  \\
  \texttt{g.roberge@elsevier.com}               \\\AND
  Jingwen Mu       \\
  Auckland University, New Zealand         \\\And
  Kevin Kang      \\
  Auckland University, New Zealand         \\\And
  Weiwei Wang        \\
  Auckland University, New Zealand         \\\AND
  Maurice Vanderfeesten         \\
  Vrije Universiteit, the Netherlands              \\\And
  Maxim Rivest \\
  Elsevier, Canada                              \\\And
  Savvas Chamezopoulos\\
  Elsevier, the Netherlands      \\\AND
  Robert Jaworek \\
  Palacký University, Czech Republic  \\\And
  Ma\'eva Vignes \\
  University of Southern Denmark \\\And
  Bamini Jayabalasingham \\
  Elsevier, the USA \\\AND
  Finne Boonen \\
  Elsevier, the Netherlands \\\And
  Chris James \\
  Elsevier, the Netherlands \\\And
  Marius Doornenbal \\
  Elsevier, the Netherlands \\\And
    Isabelle Labrosse \\
  Elsevier, Canada \\\AND
}

\begin{document}

\maketitle

\begin{abstract}
The United Nations (UN) Sustainable Development Goals (SDGs) challenge the global community to build a world where no one is left behind. Recognizing that research plays a fundamental part in supporting these goals, attempts have been made to classify research publications according to their relevance in supporting each of the UN’s SDGs. In this paper, we outline the methodology that we followed when mapping research articles to SDGs and which is adopted by Times Higher Education in their Social Impact rankings. We compare our solution with other existing queries and models mapping research papers to SDGs. We also discuss various aspects in which the methodology can be improved and generalized to other types of content apart from research articles. The results presented in this paper are the outcome of the SDG Research Mapping Initiative that was established as a partnership between the University of Southern Denmark, the Aurora European Universities Alliance (represented by Vrije Universiteit Amsterdam), the University of Auckland, and Elsevier to bring together broad expertise and share best practices on identifying research contributions to UN’s Sustainable Development Goals.


\end{abstract}

\section*{Introduction}

Numerous approaches to mapping research to the United Nations (UN) Sustainable Development Goals (SDGs)\footnote{\url{http://metadata.un.org/sdg/}} have been documented \parencite{elsevier2020_sdg_queries,bergen_SDG,bordignon_2020_sdg_mapping,confraria2021,LaFleur2019}. These approaches vary with regard to the framework used to define inclusion and exclusion criteria, the methodology employed to retrieve publications, and the publication database used. For example, the approach to defining inclusion and exclusion criteria may be set conservatively to limit publications to those documenting actions made to achieve the SDG targets, or conversely, may be set using a more liberal approach, thereby including any papers that increase knowledge on the overall topic. With regard to the methodology employed to retrieve publications, publication sets for a specific SDG can use a Boolean approach only or be complemented by machine learning algorithms. The source of publications that the methodology is applied to can also introduce variability, given the availability of many data sources, ranging from open access, subscription-based, or a mixture of both.

To date, there is no broadly agreed-upon methodology for mapping research to the SDGs, and existing methods produce quite different results \parencite{bergen_SDG}. A common approach to identifying research related to a topic is to use Boolean search expressions. The Boolean method involves the use of keywords, either alone or in combination, using conditional functions and applied to specified text sections (title, abstracts, keywords, etc.) of scientific publications, and results in the exclusive retrieval of articles within which the defined search expressions were found. The authors of \parencite{bergen_SDG} applied the Boolean method, taking an approach to limit their SDG publication sets to publications with a direct contribution to targets and/or indicators, with efforts made to reduce the impact of issues raised on the Boolean technique, resulting in a more restrictive publication set. Bordignon’s strategy \parencite{bordignon_2020_sdg_mapping} aimed at reducing the polysemy of terms by limiting keywords from Elsevier 2020 queries \parencite{elsevier2020_sdg_queries} to relevant subject areas using the All Science Journal Classification (ASJC). A text-mining tool (CorTexT) was then used to enrich those selected publications. The Aurora European Universities Alliance \parencite{schmidt_felix_2021_4917171} developed and released their 169 target-level SDG queries \parencite{vanderfeesten_maurice_2020_3817445} also using keyword combinations, boolean- and proximity operators. The University of Auckland \parencite{sdg_auckland} developed queries informed by the researchers within their network, resulting in a localized version that takes into account more papers that are specific to Australian and New Zealand research topics. In \parencite{confraria2021}, the authors employ a two-step approach involving building SDG-specific terms obtained from many sources (policy reports, publications, forums, etc), applying a selection process to the terms and then using the terms to identify citation-based communities of publications. However, as described in \parencite{bergen_SDG}, such a keyword-based-based approach involves challenges related to the interpretation of the themes and concepts of the SDGs, decisions around which publications to designate as a ``contribution'' to the chosen interpretation of the SDG, and the translation of concepts into a search query that will accurately identify publications.

An alternative or complementary approach to query-based methods involves using machine learning to map research articles to SDGs: either in a supervised manner, i.e., performing classification, or an unsupervised manner, i.e., performing clustering. Supervised methods typically resort to the same SDG queries to obtain a labeled dataset to train the model \parencite{maeva_2020,SouthAfricanSDGHub2020}. Clustering is typically done with paper text representations or citation graphs where the resulting clusters are later mapped to SDGs either directly or via intermediate clusters, e.g., ``topics'' \parencite{DigitalScience2020, clarivate_sdgs2019}. Refer to \parencite{pukelis2020osdg} for an overview of some more methods of classifying documents into SDGs. However, they all face the same challenges noted above, and machine learning further introduces the problem of interpretability of the model predictions or the clusters attained. 

Since 2018, Elsevier has endeavored to map research to the SDGs, releasing publicly available queries to facilitate transparency and reproducibility \parencite{elsevier2020_sdg_queries}. Herein, we describe the approach taken to improve former attempts to map research to the SDGs, taking feedback into account, resulting in the creation of a more comprehensive query set with sub-queries addressing targets and indicators and the application of a machine learning model to increase recall. This methodology (``Elsevier 2021 SDG mapping'' \parencite{elsevier2021_sdg_queries}) captures, on average, twice as many articles as the 2020 version while keeping precision above 80\%. Times Higher Education (THE) is using Elsevier SDG mapping as part of their Social Impact rankings \parencite{the_impact_ranking}. ``Elsevier 2023 SDG mapping'' \parencite{Els2023queries} is the most up-to-date simplified version of the queries \& ML model differing from the 2021 version in Covid-related enhancement to SDG 3 queries and queries designed for SDG 17 ``Partnerships for the goals''.

To evaluate the approach, the output generated using the developed methodology was compared to the results generated by Aurora European Universities Alliance \parencite{multilingual_sdgs_Vanderfeesten_jaworek_2022}, the University of Auckland \parencite{sdg_auckland}, the University of Bergen \parencite{bergen_SDG}, SIRIS Academic \parencite{Duran-Silva2019},
Bordignon queries \parencite{Bordignon2021}, and the ML classifier by the South-African SDG Hub \parencite{SouthAfricanSDGHub2020}.

We haven’t seen much research aimed at doing similar benchmarking of different SDG mapping approaches with hand-labeled datasets. \parencite{wulff2023} is the closest investigation to ours; apart from benchmarking, the authors explore the extent to which SDG queries produce false positives by marking non-SDG-related content with SDG labels. They also investigate the bias in SDG labeling systems defined as the normalized difference in the number of predicted and observed (i.e. put by human experts) SDG labels. 

The novel contributions of this paper can be summarized as follows: 
\begin{itemize}
    \item We solve the problem of recall assessment for keyword queries mapping research articles to Sustainable Development Goals, while other approaches typically focus on precision;
    \item  We are among the first to quantitively evaluate existing sets of such keyword queries against several validation datasets.
\end{itemize}


\section{Methodology}
\label{sec:methodology}
\subsection{Developing SDG queries}

The SDGs are goals to achieve rather than research topics, each SDG encompassing many targets. Using Boolean search expressions to build SDG-specific publication sets presents many challenges. Elsevier implemented a bottom-up approach to the construction of each SDG-relevant publication set, whereby several sub-queries were first constructed for each SDG target, and then aggregated at the SDG level.

\subsubsection{Building a query for each target within an SDG}
\label{science_metrix}

Criteria for delineating the publication sets relevant to each SDG were designed by a team consisting of a minimum of four analysts and were based on an extensive literature review done by the team to gain an understanding of the SDG. As a first step, the SDG was further subdivided into themes to facilitate the creation of specific criteria linked to specific SDG targets. The criteria defined for each theme aimed to specify topics of focus as well as any requirements for ``action terms'' in association with the topics \parencite{bergen_SDG}. For example, for the topic of ``poverty'', the action term ``alleviate'', or other action terms holding similar meaning might be deemed a requirement. To ensure homogeneity in the approach, criteria developed by the team of analysts were submitted to a review committee consisting of both those on the SDG team and those external to the team. The review committee was responsible for reviewing the criteria, recommending changes, and final approval of the criteria. Table \ref{table1} presents the criteria for SDG1 overall (SDG1-Main) and subcategories related to SDG1. These criteria were defined for each SDG and theme related to the SDG.

\begin{table}[!ht]
\caption{An example of SDG 1 subtopics and associated SDG targets.}
\centering
\begin{tabular}{|p{0.1\textwidth}|p{0.4\textwidth}|p{0.4\textwidth}|}
\hline
\textbf{Subset code} & \multicolumn{1}{c|}{\textbf{Criteria}}  & \multicolumn{1}{c|}{\textbf{Associated target}}  \\ \hline
SDG1-Main   & Research focused on poverty and research as defined for any SDG1-subset below. ``Action term'' specified: the action term, ``alleviate'' was  applied to make the topic term ``poverty'' more specific. & Target 1.1: eradicate extreme poverty \par Target 1.2: reduce poverty by half All Targets associated with SDG1-Subsets  \\ \hline
SDG1-Theme1 & Research focused on social programs, including all articles discussing social security systems related to health, finance, and work. No ``action terms'' were required for the inclusion of the topics above. & Target 1.3: Implement nationally appropriate social protection systems   \\ \hline
SDG1-Theme2 & Research focused on microfinance, access to property, inheritance, natural resources, and new technologies as they relate to facilitating access, equality, and human rights. ``Action term'' specified: the action term, ``access to'' was applied.              & Target 1.4: equal rights to economic resources and basic services        \\ \hline
SDG1-Theme3 & Research focused on resilience, exposure, and vulnerability to disasters (financial, climate-related, social..), particularly on understanding poor and vulnerable people and communities.  No ``action terms'' were required for the inclusion of the topics above. & Target 1.5: build the resilience of the poor  \\ \hline
SDG1-Theme4  & Research focused on financial aid, policies, government support (such as food banks and support distribution strategies), and strategies to eradicate poverty.  No ``action terms'' were required for the inclusion of the topics above.  & Target 1A: Ensure significant mobilization \par of resources from a variety of sources \par Target 1B: Create sound policy frameworks \\ \hline
\end{tabular}

\label{table1}
\end{table}

Following the establishment of criteria defining the research areas of focus relevant for each SDG (overall and per SDG-Theme), these criteria were used to guide the development of queries to retrieve publication sets. Where possible, the analyst responsible for query development was selected due to subject matter expertise in the field. Otherwise, the process was informed by a literature review. An iterative approach was taken to assess the precision with which individual keywords and sets of keywords identified publications that met the criteria. Keywords from the Elsevier 2020 \parencite{elsevier2020_sdg_queries} and Aurora European Universities Alliance \parencite{schmidt_felix_2021_4917171} queries were assessed first. Additional keywords were identified using term-frequency and inverse-document frequency (TF-IDF) analyses of text from titles, abstracts, and author keywords from publications meeting the criteria. Additional efforts were taken to identify publications that may have been excluded based on the developed query. Specifically, (1) the query results were analyzed to identify specialized journals that would be expected to include a high percentage of publications that fit the criteria, and (2) the citation network of the publications retrieved using the query was assessed to identify publications within the citation network of the results (i.e., publications citing or cited by the publications retrieved by the query) that were not retrieved by the query. Publications from these specialized journals or the citation network that were not being retrieved by the query were assessed to identify additional keywords to include in the query to increase recall. Relevant exclusions were built into the queries to increase precision and could result in the exclusion of specific terms using Boolean operators or the exclusion of fields of science deemed to be outside the scope of the criteria. 

To facilitate the continuous evaluation of the query, publications were manually reviewed to assess their fit against the criteria shown above (see Table \ref{table1} for SDG1). An evaluation of a minimum of 100 random publications by two independent analysts was done to support the calculation of precision metrics for each query, and a minimum precision threshold of 90\% was required for a query to be considered acceptable. The recall was assessed against independent publication sets developed by an analyst consisting of publications from specialized journals identified to fit the criteria. As most specialized journals do not exclusively focus their content on a single SDG, a minimum recall of 60\% was required for a query to be considered acceptable. In cases where no single journal was specific enough for all publications within that journal to fit the criteria set for an SDG or SDG-Theme, a publication set was constructed by manually selecting publications from a journal with high relevance to the SDG (or SDG-Theme), and recall was assessed against this set.

\subsubsection{Precision assessment}

As described earlier, queries were composed gradually, starting from the seed queries developed first by analysts. These queries were developed by concatenating queries together with Boolean ``OR'' expressions after evaluating the keywords suggested by the TF-IDF analysis on the seed dataset. Before adding a new search expression to the global SDG dataset, analysts were encouraged to sample at least 10 documents to ensure high precision was maintained throughout most queries and not simply for the global SDG dataset. This is quite important; otherwise, some keywords bringing a small number of new publications but covering mostly content not relevant to the SDG could be included in the dataset, and while their impact on global precision would be relatively small, it would still mean that analyst would be forcing bad content with such terms. The sampling was performed directly in the exploration window, which could be used to quickly draw random samples of publications containing the selected keywords. This enabled analysts to vet new keywords quickly, which was necessary given the complexity of the queries needed to delineate the SDGs properly. As a target, a 90\% precision level was required to commit the tentative search expression, as otherwise, a lower level would lead to diminished precision for the global dataset at the end of the iterative process. This was especially critical for keywords adding a lot of new documents to the global dataset as lower precision for these would more greatly influence the global precision.

Although precision was assessed throughout the whole process, a more formal precision estimate was performed at the end of the whole process to provide a final assessment, which would guide analysts as to whether they could stop their work or if an additional effort was needed to remove content that was deemed too broad and resulted in lowered precision. A large sample of 100 publications was pulled from the global dataset, and analysts performed a manual inspection of these, the tool enabling to tag publications as good, bad, and also in-between for cases where the analyst was unsure if documents should be included or not. This feature presented the advantage that final precision assessments were stored in the tool and could be consulted at any time in the future. This was especially helpful when additional validation steps were performed by the QA analyst, who was able to validate the precision assessment by assessing the same sample. If final precision was in the 90\%-95\% range or above, precision was deemed sufficient.

As a final step, a final QA was performed by an expert bibliometrician with more than a decade of experience in the field and in building datasets. Each query was analyzed by this expert and tested again for precision, reusing the samples pulled from each analyst but often pulling new samples as well to further solidify confidence. This additional layer of validation helped cement the process, ensuring a unified view of all SDGs in a similar way to what was accomplished when defining definitions as groups at the beginning of the process. The QA round led to multiple modifications, removals, and additions across most SDGs, often resulting in relatively minor changes in publication counts but further increasing the robustness of the alignment between the definitions and the final content retrieved by the queries.

\subsubsection{Recall assessment}
\label{recall}

To determine the recall of the queries developed by analysts, a selection of specialized journals was identified for each SDG to serve as a stand-in for a gold standard, representing the subjects at hand. This pragmatic ‘proxy’ for recall measurement was developed in the absence of a true gold standard for testing the recall of the queries. The absence of a gold standard is unsurprising; should such a gold standard exist, it would imply that perfectly delineated document sets for SDGs would already exist, thus rendering the current exercise irrelevant. For each SDG, sets of highly relevant journals were identified using a combination of keyword searches in journal names and percentages of journal content covered by the keyword queries. This dual approach ensured that no relevant journals would be missed simply because their name was not declarative enough to be captured. After these journals were identified, analysts aimed to maximize recall across each of these journals while maintaining high precision. Recall levels of 60\%-70\% were set as the original minimal level for the current exercise based on two decades of expertise in building such datasets. Increasing recall for some categories without comprising precision is sometimes easy in subjects relying on highly declarative vocabulary, while it can become quite tricky in others, especially those mixing multiple dimensions as their core concepts. In the case of the targets of the SDGs, this notion is especially relevant, as SDGs often mix basic research with economic and social concepts.

During the process, recall against the selected gold standard of journals was tested frequently to determine if more investigation was needed to add new keywords to the queries. Analysts performed recurring analyses of the content of these journals not captured by the queries to detect any research subject not covered. TF-IDF analyses on these documents not retrieved were performed to obtain lists of suggested terms for inclusion to further increase recall. At the end of the process, if recall remained low, corrected recalls were computed by sampling amongst the publications not retrieved with the keyword queries, estimating which part was truly relevant to the subject at hand. Indeed, specialized journals, while usually having targeted scopes, are not always fully relevant to the topic at stake. By sampling about 50 publications, analysts were able to compute corrected recall scores by estimating the fraction of the content not covered that was indeed relevant to subjects.
 
As a final step, a final QA was again performed by the expert bibliometrician. Each query was analyzed by this expert and tested for recall, investigating whether areas of each target might have been missed or left out by the analyst. The QA round led to multiple modifications, removals, and additions across most SDGs, often resulting in relatively minor changes in publication counts but further increasing the robustness of the alignment between the definitions and the final content retrieved by the queries.

Below, we refer to the mentioned recall evaluation dataset as to the \textit{Elsevier recall dataset}.

\subsection{Machine learning applied to SDG classification}

On top of the mapping produced by the queries described above, additional articles are mapped to the SDGs by a machine learning model.  
 
In a nutshell, the model is a logistic regression trained with TF-IDF representations of titles, keywords, abstracts, and two more optional text fields -- main terms extracted from the full text and subject areas of the journal that published the paper. Thus, the model learns similar keyphrases for each SDG and helps to improve the recall of the queries. To keep precision high, we keep only those papers that are classified by the model with 95\% or higher predicted probability for some SDG. 
 
In the ``Elsevier 2021 SDG mapping'' release \parencite{elsevier2021_sdg_queries}, the Elsevier team specifies the input data for the model, the targets that it’s trained with, the technical details of the model itself, and model performance. Also, to ease the interpretation of the model classification outcomes, we share the SDG-specific key phrases learned by the model, as well as sample articles classified by the model. Please refer to the mentioned documentation for more details on the machine learning component of our approach. 

\subsection{Combining the queries and the model}

\begin{figure}[!ht]
\caption{Distribution of the number of papers mapped by the queries (SM, orange) and by the model (ML, blue), by SDG (ignoring SDG 17).}
\centering
\includegraphics[width=0.7\linewidth]{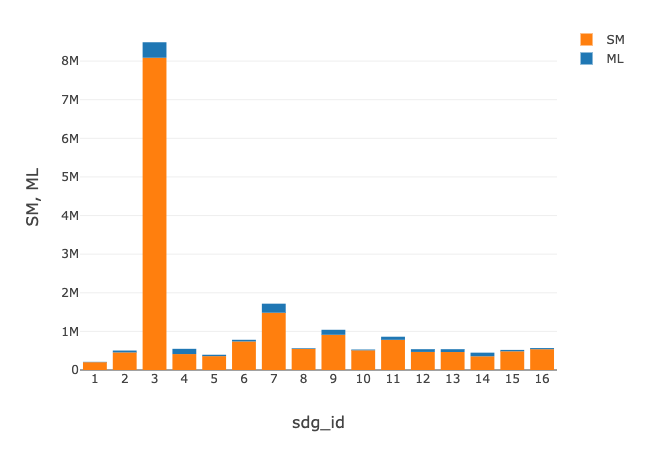}
\label{fig:sdg_dist}
\end{figure}

The end-to-end approach to mapping scholarly records to SDGs is two-staged:		
\begin{itemize}
    \item first, the keyword SDG queries are run (orange in  Fig. \ref{fig:sdg_dist})
    \item then, the ML model adds about 3.5\% of papers (blue in  Fig. \ref{fig:sdg_dist}) on top of what is classified by the keyword queries. We only keep the most confident model predictions by thresholding predicted scores at 0.95. 
\end{itemize}

It’s worth noting that the approach is limited to the Scopus database as the queries are written in Scopus search syntax.

\section{Results}
\label{sec:results}

\subsection{Comparison between the SDG queries}

Below, we describe the SDG queries and validation datasets that we used for the comparison in terms of precision, recall, and F1 scores.

\subsubsection{Query models}

Table \ref{table2} describes the different classification methods that we compared. These can be either keyword queries (``Elsevier queries 2020'', ``Aurora queries v5'', ``Auckland queries v2'', and ``Bergen SDG queries'') or machine learning models (``Aurora ML v0.2'') or both (``Elsevier queries+ML 2021'', ``Elsevier queries+ML 2022'').

\begin{table}[!ht]
\caption{SDG classification methods (both keyword queries and ML models) used in the evaluation.}
\centering
\begin{tabular}{|p{0.25\textwidth}|p{0.7\textwidth}|p{0.05\textwidth}|}
\hline
\textbf{Classification method }                                                                                   & \textbf{Description }                                                                                                  & \textbf{Web location  }                                                \\ \hline
Auckland queries v2 \par \textbf{(Auckland\_v2)    }                                                                & To gain a better understanding of our research contribution, the University of Auckland SDG Keywords Dictionary Project seeks to build on the processes developed by the United Nations and THE in order to create an expanded list of keywords that can be used to identify SDG-relevant research       \parencite{Wang2023}  & \href{https://www.sdgmapping.auckland.ac.nz/}{link}                        \\ \hline
Aurora ML v0.2 \par   \textbf{(Aurora\_ml)} & ``AI for mapping multi-lingual academic papers to the United Nations’ Sustainable Development Goals (SDGs)'' \parencite{multilingual_sdgs_Vanderfeesten_jaworek_2022}  & \href{https://doi.org/10.5281/zenodo.5603019}{link}                        \\ \hline
Aurora queries v5 \par \textbf{(Aurora\_v5)    }                       & “Mapping Research Output to the Sustainable Development Goals (SDGs)” \parencite{vanderfeesten_maurice_2020_3817445}                                       & \href{https://zenodo.org/records/4883250}{link}                            \\ \hline
Bergen SDG queries \par \textbf{(Bergen\_2023\_baa}  \par \textbf{and Bergen\_2023\_bta)} & The Bergen approach created queries for Web of Science to retrieve SDG-related publications for a limited number of SDGs. The queries have been translated for Scopus and a sample of the results has been taken as positive examples. These have been supplemented by other publications which did not appear in the queries as negative examples. Two datasets were created, one based on the Action Approach queries and one based on the Topic Approach queries – referred to as Bergen TAA and Bergen TBA respectively \parencite{bergen_SDG} & \href{https://zenodo.org/records/7711561}{link}                            \\ \hline
Elsevier queries 2020\par \textbf{(Els\_2020) }                        & “Identifying research supporting the United Nations Sustainable Development Goals” \parencite{elsevier2020_sdg_queries}                                                                                                                   & \href{https://elsevier.digitalcommonsdata.com/datasets/87txkw7khs/1}{link} \\ \hline
Elsevier queries+ML 2021 \par \textbf{(Els\_2021)  }                  & ``Improving the Scopus and Aurora queries to identify research that supports the United Nations Sustainable Development Goals (SDGs) 2021'' \parencite{elsevier2021_sdg_queries}   & \href{https://elsevier.digitalcommonsdata.com/datasets/9sxdykm8s4/4}{link} \\ \hline
Elsevier queries+ML 2022 \par (Els\_2022)                         & A simplified version of ``Elsevier queries+ML 2021'' with Covid-related addendum to SDG 3 \parencite{elsevier2022_sdg_queries}  & \href{https://elsevier.digitalcommonsdata.com/datasets/6bjy52jkm9/1}{link} \\ \hline
Elsevier queries+ML 2023 \par \textbf{(Els\_2023)  }                     & For 2023, the SDGs use the exact same search query and ML algorithm as the Elsevier 2022 SDG mappings, with only minor modifications to five SDGs, namely SDG 1, 4, 5, 7 and 14. In these cases, the queries were shortened by removing exclusion lists based on journal identifiers. These exclusion lists often contained thousands of items to filter out content in journals that were not core to the SDGs. \parencite{Els2023queries} & \href{https://elsevier.digitalcommonsdata.com/datasets/y2zyy9vwzy/1}{link} \\ \hline
South African SDG hub \par \textbf{(South\_africa)}  & A Machine Learning model mapping text to SDGs  & \href{https://sasdghub.up.ac.za/home/}{link}                             \\ \hline
SIRIS SDG queries \par \textbf{(SIRIS)  }  & The SIRIS queries were developed by extracting key terms from the UN official list of goals, targets, and indicators as well as from relevant literature around SDGs. The query system has subsequently been expanded with a pre-trained word2vec model and an algorithm that selects related words from Wikipedia. There are multiple queries per SDG \parencite{Duran-Silva2019} & \href{https://zenodo.org/records/4118028}{link}                            \\ \hline
Bordignon SDG queries \par \textbf{(Bordignon) }                           & These queries aimed at reducing the polysemy of terms by limiting keywords from Elsevier 2020 queries \parencite{elsevier2020_sdg_queries} to relevant subject areas using the All Science Journal Classification (ASJC) \parencite{bordignon_2020_sdg_mapping}   & \href{https://data.mendeley.com/datasets/xrx7ddbbb4/1}{link}               \\ \hline
\end{tabular}
\label{table2}
\end{table}

\subsubsection{Validations sets; collection method, sizes, and quality}

\begin{table}
\caption{SDG validation datasets used in the evaluation.}
\centering
\begin{tabular}{|p{0.15\textwidth}|p{0.4\textwidth}|p{0.06\textwidth}|p{0.06\textwidth}|p{0.25\textwidth}|}
\hline
\textbf{Validation set   }                                                                    & \textbf{Description \& method of collection} & \textbf{Web location}                                     & \textbf{Size}  & \textbf{Remarks on quality }    \\ \hline
Elsevier recall dataset\par \textbf{(Elsevier\_recall)} & See Section \ref{recall}    & Shared via ICSR Lab, see Sec. \ref{sec:data}                   & 465k  & The dataset is noisy in the sense that not all papers from an SDG-specific journal are relevant to the same SDGs. Hence, we don’t aim at 100\% recall with respect to this dataset                                             \\ \hline
Aurora Survey \par \textbf{(Aurora1)  }             & ``Survey data of "Mapping Research output to the SDGs" by Aurora European Universities Alliance (AUR)'' 244 senior researchers from different universities in Europe and the US filled in a survey. They were only allowed to enter the survey if they were familiar with the SDG they had selected to evaluate. The first question was to provide a list of research papers they believe are relevant to that selected SDG. The second question was to handpick from a given set of 100 randomly drawn papers in the Aurora query result set, the papers they believe (based on reading the title, abstract, journal name, and authors) belong to the selected SDG. The suggested papers and the selected papers are included in the validation set. \parencite{aurora_sdg_survey2020} & \href{https://zenodo.org/records/3813230\#.YyG93uxBxYw}{link} & 6741  & Bias: the researchers are located at Western European universities. \\ \hline
Aurora Suggested \par Papers \par \textbf{(Aurora2) }  & The papers suggested by researchers, see ``Survey data of "Mapping Research output to the SDGs"''. & \href{https://zenodo.org/records/3813230\#.YyG93uxBxYw}{link} & 3964  & The researchers involved in the survey identified themselves as having expertise in a specific SDG. They might also have the incentive to cite their own research. \\ \hline
Elsevier multi-label SDG dataset \par \textbf{(Els\_multilabel)  }   & The dataset consists of 6000 papers annotated by 3 experts each. These papers come from 5 data sources to span as diverse as possible set of SDG-related papers. 30\% of the papers are not mapped to any of SDGs & Shared via ICSR Lab, see Sec. \ref{sec:data}. & 6000  & Annotators are not as versed in SDGs as the analysts who developed Elsevier queries and the Elsevier recall dataset  \\ \hline
Chilean \par multi-label \par \textbf{(Chile)  }                                                        & The dataset is provided by Pontificia Universidad Católica (PUC) based in Chile and consists of about 1200 papers self-assessed by PUC researchers and labeled with 0, 1, or 2 SDGs. & \href{https://repositorio.uc.cl/handle/11534/61951}{link}     & 1200  & Biases: self-assessment, only Chilean researchers \\ \hline
OSDG Community Dataset \par  \textbf{(OSDG)}          & A public dataset of thousands of text excerpts, which were validated by approximately 1,000 OSDG Community Platform (OSDG-CP) citizen scientists from over 110 countries, with respect to the Sustainable Development Goals. \parencite{pukelis2020osdg}  & \href{https://zenodo.org/records/6831287\#.YyMF5OxBxYy}{link} & 32431 & Crowd-sourced dataset, the annotators are not versed in SDGs. In our benchmarks, we only kept the records with the difference between positive and negative votes greater than or equal to 2. Thus leaving only 26217 records. \\ \hline
\end{tabular}

\label{table3}
\end{table}

Table \ref{table3} provides details on the validation datasets used in the comparison. It also mentions the associated limitations and biases. It is important to mention that there’s no single best validation dataset to evaluate the output of SDG classification. 

\subsubsection{Performance; query models measured against validation sets}

Table \ref{table:results_f1_scores} provides the evaluation results for the SDG classification methods outlined in Table \ref{table2} and evaluation datasets described in Table \ref{table3}. Each cell shows 2 values: micro-average F1-score and macro-average F1-score (the micro-average F1-score aggregates performance metrics across all classes by treating each instance equally, while the macro-average F1-score computes the F1-score for each class independently and then takes the average, giving equal weight to all classes regardless of their sizes), in percent (\%). Both precision and recall were calculated with respect to the validation sets, i.e., all predictions beyond the validation sets were ignored:

\begin{itemize}
    \item precision is calculated as the number of correctly predicted SDG IDs divided by the number of Scopus IDs tagged with the same SDG ID in the given validation set;
    \item recall is calculated as the proportion of correctly predicted SDG IDs within the given validation set.
\end{itemize}

To compare with Bergen queries, Table \ref{table:results_f1_scores_6sdgs} provides similar metrics only considering a subset of 10 SDGs, namely, SDG 1 (No poverty), SDG 2 (Zero hunger), SDG 3 (Good health and well-being), SDG 4 (Quality education), SDG 7 (Affordable and clean energy), SDG 11 (Sustainable cities and communities), SDG 12 (Responsible consumption and production), SDG 13 (Climate action), and SDG 14 (Life below water), and SDG 15 (Life on land).

The same comparisons for precision and recall are found in the Appendix, see. Tables 9-12. 

Note that micro-averaging favors well-represented, frequent classes (like SDG 3 in our case), while high macro-averaged scores mean that the method works fairly well across all SDGs because bad results for a single SDG affect macro-averaged metrics much more than micro-averaged ones. By attending to both micro- and macro-averaged F1 scores, we try to assess both aspects: how well the method is at classifying papers into frequent or rare classes.

The code reproducing the experiments presented in this subsection is found on GitHub\footnote{\url{https://github.com/Yorko/sdg_mapping_queries_n_ml_benchmarks}}. Refer to Sec. \ref{sec:data} for instructions on getting data should you wish to reproduce the presented experiments.

Note that micro-averaging favors well-represented, frequent classes (like SDG 3 in our case), while high macro-averaged scores mean that the method works fairly well across all SDGs because bad results for a single SDG affect macro-averaged metrics much more than micro-averaged ones. By attending to both micro- and macro-averaged F1 scores, we try to assess both aspects: how well the method is at classifying papers into frequent or rare classes.

\begin{table}[]
\caption{F1 scores with micro/macro averaging (percentages, \%) for 10 classification methods and 5 validation datasets. Bolded is the best result in the column, asterisk for multiple ``winners'' depending on micro- or macro-averaging.
}
\centering
\begin{tabular}{|l|c|c|c|c|c|}
\hline
\textbf{Method \textbackslash\ ~dataset} & \textbf{Aurora1} & \textbf{Aurora2} & \textbf{Els\_multilabel} & \textbf{Chile} & \textbf{OSDG} \\ \hline
\textbf{Auckland\_v2} & 49/40 & 46/33 & 69/62 & 60/37 & 47/40 \\\hline
\textbf{Aurora\_ml} & 53/44* & 39/32 & 64/57 & 55/38 & \textbf{53/46} \\\hline
\textbf{Aurora\_v5} & 55/42 & 15/18 & 37/38 & 12/14 & 26/20 \\\hline
\textbf{Els\_2020} & 47/35 & 46/28 & 63/47 & 55/25 & 33/27 \\\hline
\textbf{Els\_2021} & 46/39 & 38/32 & \textbf{73/67 }& 46/34 & 41/35 \\\hline
\textbf{Els\_2022} & 46/39 & 38/32 & \textbf{73/67} & 46/34 & 41/35 \\\hline
\textbf{Els\_2023} & 45/38 & 37/30 & 72/66 & 46/31 & 42/36 \\\hline
\textbf{South\_africa} & 51/40 & 45/35* & 72/60 & \textbf{65/41} & N/A \\\hline
\textbf{SIRIS} & 36/33 & 29/25 & 49/45 & 37/30 & 37/37 \\\hline
\textbf{Bordingon} & 45/34 & 50/30* & 60/48 & 61/32 & N/A \\\hline
\end{tabular}
\label{table:results_f1_scores}
\end{table}

\begin{table}[]
\caption{F1 scores with micro/macro averaging (percentages, \%) for 12 classification methods and 5 validation datasets. Here, the validation is performed only against a subset of SDGs for which we have Bergen 2023 queries: 1, 2, 3, 4, 5, 11, 12, 13, 14, and 15.}
\centering
\begin{tabular}{|l|c|c|c|c|c|}
\hline
\textbf{Method \textbackslash\ ~dataset}  & \textbf{Aurora1} & \textbf{Aurora2} & \textbf{Els\_multilabel} & \textbf{Chile} & \textbf{OSDG} \\ \hline
\textbf{Auckland\_v2} & 59/47 & 60/42* & 78/70 & 69/46 & 59/57 \\\hline
\textbf{Aurora\_ml} & 61/48 & 47/37 & 72/63 & 66/48 & \textbf{65/62} \\\hline
\textbf{Aurora\_v5} &\textbf{ 64/50} & 17/23 & 40/46 & 13/18 & 29/27 \\\hline
\textbf{Bergen\_2023\_baa} & 15/11 & 15/11 & 15/13 & 14/12 & N/A \\\hline
\textbf{Bergen\_2023\_bta} & 17/16 & 17/15 & 22/25 & 16/14 & N/A \\\hline
\textbf{Els\_2020} & 54/39 & 61/37 & 71/52 & 63/33 & 39/37 \\\hline
\textbf{Els\_2021} & 55/45 & 46/38 & \textbf{80/74} & 53/42 & 47/44 \\\hline
\textbf{Els\_2022} & 55/45 & 46/38 & \textbf{80/74} & 53/42 & 48/45 \\\hline
\textbf{Els\_2023} & 54/43 & 46/37 & 80/73 & 52/38 & 48/46 \\\hline
\textbf{South\_africa} & 60/50 & 58/45 & 80/72 & \textbf{74/56} & N/A \\\hline
\textbf{SIRIS} & 48/40 & 42/33 & 62/54 & 48/39 & 51/49 \\\hline
\textbf{Bordingon} & 53/34 & 67/40* & 68/52 & 72/40 & N/A \\\hline
\end{tabular}
\label{table:results_f1_scores_6sdgs}
\end{table}

We conclude that there is no single best approach performing well across all validation datasets: some approaches are, on average, better at precision (e.g., Elsevier 2020 and South African SDG ML model, see Tables \ref{table9} and \ref{table10}), others shine at recall (e.g., Auckland queries and Aurora ML model, see Tables \ref{table11} and \ref{table12}). This finding supports the general criticism that SDG classification faces: different mapping methods typically kick off with the same keywords but then result in poorly overlapping mappings \parencite{bergen_SDG,sdg13_comparsion}. Apart from these “qualitative” problems with SDG mappings, we now establish the ``quantitative'' problem: when evaluated against several hand-labeled SDG datasets, different approaches fail to select a clear winner.
 
We notice a clear ``overfitting'' phenomenon: Elsevier queries+ML 2022 are best when validated against Elsevier’s multi-labeled dataset, while Aurora queries v.5/Aurora ML model achieve the highest F1 scores against the Aurora survey dataset. A probable explanation is that the datasets were crafted for a specific definition/operationalization of SDGs, and these definitions are undoubtedly different from one project to another.
 
It is important to conclude that there is no single “golden” SDG validation dataset; each one considered in our experiments comes with its own shortcomings (see Table \ref{table3}, remarks on quality), and each dataset used in query development reflects some certain interpretation of SDGs by the query developers. Similarly, to how \parencite{bergen_SDG} concluded that there’s a poor overlap in publications found by different sets of queries, we conclude that there’s no clear winner among SDG classification methods when those are validated with available human-annotated SDG datasets.  

\subsection{Tracking the progress of Elsevier queries}
\label{els_progress}

The progress with SDG query development at Elsevier was tracked both in terms of recall as described in Section 1.1.3 and in terms of precision/recall/F1 when validated with the independently labeled Elsevier multi-label dataset. 

Table \ref{table_recall_dataset} shows recall scores for different Elsevier queries as measured against the Elsevier recall dataset described in detail in Section \ref{sec:methodology}. Table \ref{table:results_els_queries} shows precision, recall, and F1  scores for different Elsevier queries as measured against the Elsevier multi-label SDG dataset described in Table \ref{table3}. Note that due to the specifics of the SDG query creation methodology, it makes sense to report only recall for the first dataset. The reason is that it’s labeled in a noisy way (the assumption that all papers from an SDG-specific journal contribute to the same Goal is far from perfect); thus, looking at precision (and hence F1) is not meaningful. However, reporting recall makes perfect sense – it shows how many SDG-related papers from this large dataset the queries can detect. 

\begin{table}[]
\centering
\caption{Elsevier queries validated against the Elsevier recall dataset (see Section \ref{sec:methodology}). Micro- and macro-averaged values for recall are reported.
}
\begin{tabular}{|l|c|}
\hline
\multirow{2}{*}{\textbf{Method \textbackslash~ dataset}} & \multirow{2}{*}{\textbf{Elsevier recall dataset, recall}} \\
                                                           &                                                           \\ \hline
\textbf{Els\_2020}                                         & 54/38                                                     \\ \hline
\textbf{Els\_2021}                                         & 78/72                                                     \\ \hline
\textbf{Els\_2022}                                         & 78/72                                                     \\ \hline
\textbf{Els\_2023}                                         & 73/68                                \\ \hline
\end{tabular}
\label{table_recall_dataset}
\end{table}

\begin{table}[]
\caption{Elsevier queries validated against the Elsevier multi-label SDG dataset (See Table \ref{table3}). P stands for precision, R -- for recall, F1 -- for F1-score. Micro- and macro-averaged values are reported.}
\centering
\begin{tabular}{|l|ccc|}
\hline
\multirow{2}{*}{\textbf{Method \textbackslash~dataset}} & \multicolumn{3}{c|}{\textbf{Elsevier\_multilabel}}                                             \\ \cline{2-4} 
                                                                       & \multicolumn{1}{c|}{\textbf{P}} & \multicolumn{1}{c|}{\textbf{R}} & \textbf{F1}                \\ \hline
\textbf{Els\_2020}                                                     & \multicolumn{1}{c|}{72/62}      & \multicolumn{1}{c|}{57/42}      & 63/45                      \\ \hline
\textbf{Els\_2021}                                                     & \multicolumn{1}{c|}{69/63}      & \multicolumn{1}{c|}{78/75}      & 73/63                      \\ \hline
\textbf{Els\_2022}                                                     & \multicolumn{1}{c|}{69/63}      & \multicolumn{1}{c|}{78/75}      & 73/63                      \\ \hline
\textbf{Els\_2023}                                                     & \multicolumn{1}{l|}{68/62}      & \multicolumn{1}{l|}{76/73}      & \multicolumn{1}{l|}{72/62} \\ \hline
\end{tabular}
\label{table:results_els_queries}
\end{table}

From Tables \ref{table_recall_dataset} and \ref{table:results_els_queries}, we see that all of Elsevier 20211-2023 queries perform about the same in terms of metrics and provide a considerable improvement in recall (and hence F1) over the earlier 2020 version of the queries.

The metrics are pretty close for the 2021-2023 versions of the queries because the 2022 and 2023 updates were not as considerable as the ones in 2021. Namely, the 2022 version (Roberge et al. 2022) introduced only COVID-related changes to SDG 3. The 2023 version of the queries (Bedard-Vallee et al. 2023) introduced changes to SDGs 1, 4, 5, and 14, removing long lists of journal identifiers and replacing them with keywords.

\section{Discussion and Perspectives}
\label{sec:discussion}

In previous sections, we described the methodology and evaluation results. Below, we outline possible improvements to the SDG mapping approach, including localization of the SDG queries, query generalization to non-English languages, and extending the approach to non-article content.

\subsection{Localization}
\label{subsec:localization}

Research activities do not stand alone; they are an integral part of the geographical place they were initiated and the communities they serve. An attempt to measure SDG-related research activities can be improved by infusing the local context within which the research activities take place. A localization approach can further foster understanding of, for example, the degree to which the prevailing SDG mapping approaches capture SDG research in the geographical region that may or may not have been described by keywords and keyphrases with close semantic relatedness to the keyword-based queries, e.g., \textit{Elsevier 2020 queries}.

The University of Auckland's approach \parencite{Wang2023} is one such localization attempt based on Elsevier's earlier 2020 queries, a mixture of the UN official targets and indicators, and the suggested search terms by the Sustainable Development Solutions Network (SDSN). The n-gram model was applied to two samples of Scopus publication metadata, i.e., a global publication sample and a University of Auckland publication sample. The n-gram tokens were scored by a range of factors, including counts and measures of frequency, and were then ranked by those scores. Keywords with a high rank were then evaluated in more detail and manually reviewed and improved for SDG alignments. Table \ref{table:auckland_vs_els2020} shows the number of University of Auckland publications between 2009 and 2020 captured by the University's queries compared with those captured by \textit{Elsevier 2020 queries}.

\begin{table}
\caption{Comparison of Auckland v2 queries and Elsevier 2020 queries.}
\centering
\begin{tabular}{|c|c|c|c|}
\hline
\multicolumn{1}{|l|}{\textbf{SDG}} & \textbf{Auckland queries output} & \textbf{Elsevier 2021 output} & \multicolumn{1}{l|}{\textbf{Intersection}} \\ \hline
1                                  & 522                              & 229                           & 125                                        \\ \hline
2                                  & 1975                             & 420                           & 264                                        \\ \hline
3                                  & 16894                            & 7966                          & 6894                                       \\ \hline
4                                  & 2484                             & 1043                          & 745                                        \\ \hline
5                                  & 611                              & 609                           & 360                                        \\ \hline
6                                  & 684                              & 486                           & 362                                        \\ \hline
7                                  & 1152                             & 1187                          & 799                                        \\ \hline
8                                  & 428                              & 440                           & 154                                        \\ \hline
9                                  & 1044                             & 1139                          & 519                                        \\ \hline
10                                 & 1528                             & 977                           & 500                                        \\ \hline
11                                 & 1886                             & 1462                          & 779                                        \\ \hline
12                                 & 921                              & 438                           & 158                                        \\ \hline
13                                 & 1032                             & 577                           & 466                                        \\ \hline
14                                 & 1390                             & 744                           & 552                                        \\ \hline
15                                 & 1641                             & 769                           & 473                                        \\ \hline
16                                 & 891                              & 779                           & 409                                        \\ \hline
\end{tabular}
\label{table:auckland_vs_els2020}
\end{table}

For 13 out of the 16 SDGs documented in Table \ref{table:auckland_vs_els2020}, the Auckland queries capture more SDG-related publications. In some cases, the number of publications captured by the Auckland queries doubled that captured by the Elsevier 2021 approach. A significant proportion of the additional publications are captured through localized keywords and search terms. For example, “Te Whariki”¯ – the New Zealand national curriculum document for early childhood education – was used as an SDG4 keyphrase under the Auckland approach as it pinpoints what makes a quality early childhood education curriculum with an indigenous Māori lens. It retrieved 19 SDG4 papers published by the University of Auckland, of which only 6 papers were counted by the Elsevier 2020 approach. A manual inspection of these 19 ’Te Whariki’ papers unsurprisingly suggests the high relevance of all 19 papers to SDG4 Target 4.2 on ensuring quality early childhood development, care, and pre-primary education. In some other cases, the Auckland queries also gave rise to additional keywords potentially fitting for the global settings. For example, “marine biodiversity” as an Auckland keyphrase retrieved 24 SDG14 papers published by the University of Auckland, of which 19 papers were counted by the Elsevier 2020 approach.

\begin{figure}[ht]
\caption{F1 scores for the Auckland approach applied to the Aurora, Elsevier, Chilean, and OSDG datasets.}
\centering
\includegraphics[width=0.7\linewidth]{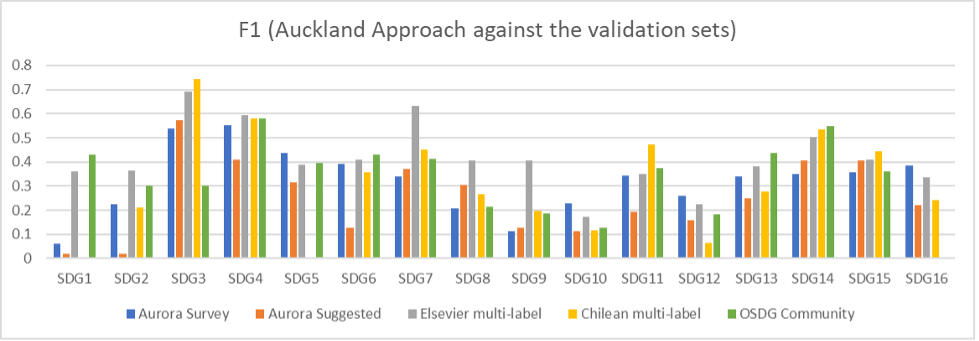}
\label{fig:sdg_auckland}
\end{figure}

As shown in Figure \ref{fig:sdg_auckland}, applying the Auckland approach to the Aurora (survey \& suggested), Elsevier (multi-label), Chilean, and the OSDG datasets generates F1 scores that are better for some SDGs (e.g., SDG3, 4, 7, and 14) than others. The F1 scores are notably low for SDGs such as SDG 1, 2, 10, and 12. 

This suggests that, while the localized approach adds useful keywords and themes in some context, further work is required to examine each keyword and keyphrase independently to understand their impact on precision and recall and to refine the search conditions upon which they should be applied. In future work, it would also be interesting to develop a contextualized SDG-label set that aligns with the contextualized SDG mapping approach (e.g., an Auckland SDG validation set) to better test out the performance of the contextualized approach against more generic, global approaches.

\subsection{Multilingual queries}

In CRISs systems\footnote{\url{https://en.wikipedia.org/wiki/Current_research_information_system}} and repositories, there are many more publications that are not included in Scopus and are written in the local language of the country to serve a different audience. We found out we could not simply replace the keywords in the queries and have the search work the same in other languages because of the syntax and morphology rules. That's why Aurora chose to train mBERT models to classify SDGs. Due to the lack of non-English SDG labeled data, we used only English training data, specifically paper abstracts.

During the evaluation, the models for SDGs 1 to 5 and 11 were applied to classify 888 German paper titles. To have a qualitative benchmark, we performed a manual SDG classification only on titles as well. In doing the latter, we tended to take a strict approach and tried to stick very close to the respective SDG indicators (e.g., non-assignment of SDG4 to publications on teacher training in Germany, as the SDG indicators only refer to teacher training in the Global
South).

The manual classification resulted in 43 SDG-related publications, whilst the ML models resulted in 58 publications SDG-related publications. The total overlap between these two methods was 8 publications. It was mainly for SDG3 – Good Health and Well-being (5/8) and can most likely be explained by great similarities in terminology between English and German for issues such as multiple sclerosis, psychotherapy, suicide, alcohol, and illegal drugs (in German: Multiple Sklerose, Psychotherapie, Suizid, Alkohol, illegale Drogen).

At the current phase of evaluation, the multilingual ability of the ML models for research output in German cannot be positively assessed. However, further analysis, including the abstracts of publications for the classification of the ML models, may offer improvements in classification quality.

\subsection{Generalization to other types of content}

In addition to SDG-related research outputs, higher education institutions have a strong interest in understanding SDG-related educational activities, as done in the Aurora SDG Course Catalogue.\footnote{\url{https://bit.ly/aurora-sdg-courses}} These SDG labels have been added manually by the course coordinators, but such a process is labor-intensive and not sustainable since this needs to be done year after year. 

Similar to publication metadata (e.g., title, abstract, keywords), many course catalogs and curriculum management systems capture metadata in a similar way (e.g., title, course short description, course long description). Whether the SDG research mapping techniques can be translated and applied to SDG course mapping represents an interesting topic to many. 

A study was conducted by the University of Auckland to apply the Auckland queries to classify courses taught by the university. The mapping results identified 792 SDG courses out of 2441 courses in total offered to students in the academic year 2020. Compared with the frequency and distribution of keywords in research mapping, course mapping demonstrated a higher concentration of keywords used to convey the SDG topics. For example, the 24 University of Auckland courses related to SDG14 are fully captured by the top ten keywords in the Auckland queries by frequency (i.e., marine; fisheries; coastal management; pollut*; aquaculture; marine environment; fisheries management; eutrophical*; aquatic ecosystem; alga*).

\section*{Conclusion}
\label{sec:conslusion}

In this paper, we outlined the methodology behind research mapping to the United Nations (UN) Sustainable Development Goals (SDGs), how it compares to other existing methods, and how well it performs with existing SDG validation datasets. We conclude that there is no single best approach performing well across all validation datasets, although Elsevier queries are slightly more stable. We also conclude that there's no single ``golden'' SDG validation dataset; each one considered in our experiments comes with its own shortcomings, and each dataset used in query development bears the intrinsic bias of the SDG interpretations by the query developers. We observed that Elsevier's queries have seen a measurable improvement from the original 2020 version to the 2021/2022/2023 versions. Finally, we discussed possible improvements to the existing approach: localization of the queries and generalization to other languages and data types. 

\section*{Data availability}
\label{sec:data}
The data underlying the results presented in the study (including the processed version of publicly available datasets listed in Table \ref{table3}) are available for scholarly research and upon \href{https://www.elsevier.com/icsr/icsrlab/how-to-apply}{application}, from Elsevier BV on the \href{https://www.elsevier.com/icsr/icsrlab}{ICSR Lab}. ICSR Lab is intended for scholarly research only and is a cloud-based computational platform that enables researchers to analyze large structured datasets, including aggregated data from Scopus author profiles, PlumX Metrics, SciVal Topics, and \href{https://www.elsevier.com/connect/new-dataset-offers-unique-insights-into-peer-review}{Peer Review Workbench}.

\section*{Acknowledgments}
\label{sec:acknowledgments}

This work is partly an outcome of SDG Research Mapping Initiative\footnote{\url{https://www.elsevier.com/about/partnerships/sdg-research-mapping-initiative}} that Elsevier initiated with the Aurora European Universities Alliance, the University of Auckland, and the University of Southern Denmark. We are also grateful to Scopus for providing data for the analysis. 

\newpage
\printbibliography
\newpage

\section*{Appendix}

\begin{table}[!ht]
\caption{Precision scores with micro/macro averaging (percentages, \%) for 10 classification methods and 5 validation datasets. Bolded is the best result in the column, asterisk for multiple ``winners'' depending on micro- or macro-averaging.}
\centering
\begin{tabular}{|l|c|c|c|c|c|}
\hline
\textbf{Method \textbackslash~ dataset} & \textbf{Aurora1} & \textbf{Aurora2} & \textbf{Els\_multilabel} & \textbf{Chile} & \textbf{OSDG} \\ \hline
\textbf{Auckland\_v2} & 40/33 & 40/31 & 58/52 & 60/36 & 44/44 \\\hline
\textbf{Aurora\_ml} & 48/41 & 34/29 & 59/52 & 50/35 & 45/47 \\\hline
\textbf{Aurora\_v5} & \textbf{65/52} & 32/38 & 63/60 & 46/40 & 51/47 \\\hline
\textbf{Els\_2020} & 52/42 & \textbf{59/40} & 72/62 & 79/42* & \textbf{51/49} \\\hline
\textbf{Els\_2021} & 44/38 & 42/37 & 69/63 & 67/46 & 41/44 \\\hline
\textbf{Els\_2022} & 44/38 & 42/37 & 69/63 & 67/46 & 42/44 \\\hline
\textbf{Els\_2023} & 43/37 & 42/36 & 68/62 & 68/44 & 43/44 \\\hline
\textbf{South\_africa} & 52/43 & 44/37 & \textbf{74/65} & 71/54* & N/A \\\hline
\textbf{SIRIS} & 25/25 & 22/21 & 36/35 & 31/27 & 27/32 \\\hline
\textbf{Bordingon} & 41/30 & 47/29 & 57/46 & 62/32 & N/A \\\hline
\end{tabular}
\label{table9}
\end{table}

\begin{table}[!ht]
\caption{Precision scores with micro/macro averaging (percentages, \%) for 12 classification methods and 5 validation datasets. Here, the validation is performed only against a subset of SDGs for which we have Bergen 2023 queries: 1, 2, 3, 4, 5, 11, 12, 13, 14, and 15.
}
\centering
\begin{tabular}{|l|c|c|c|c|c|}
\hline
\textbf{Method \textbackslash~ dataset} & \textbf{Aurora1} & \textbf{Aurora2} & \textbf{Els\_multilabel} & \textbf{Chile} & \textbf{OSDG} \\ \hline
\textbf{Auckland\_v2} & 53/43 & 53/38 & 70/62 & 71/47 & 64/62 \\\hline
\textbf{Aurora\_ml} & 65/52 & 47/38 & 73/63 & 67/48 & 69/66 \\\hline
\textbf{Aurora\_v5} & \textbf{74/60} & 36/48 & 71/71 & 52/48 & 65/65 \\\hline
\textbf{Bergen\_2023\_baa} & 47/56 & 51/20 & 51/41 & 49/10 & N/A \\\hline
\textbf{Bergen\_2023\_bta} & 32/39 & 28/28 & 43/53 & 37/32 & N/A \\\hline
\textbf{Els\_2020} & 65/52 & \textbf{71/49} & 82/72 & 86/55* & \textbf{70/67} \\\hline
\textbf{Els\_2021} & 62/51 & 57/47 & 82/76 & 81/59 & 63/61 \\\hline
\textbf{Els\_2022} & 62/51 & 57/47 & \textbf{83/76} & 81/60* & 63/61 \\\hline
\textbf{Els\_2023} & 62/50 & 58/45 & 82/75 & 80/57 & 63/62 \\\hline
\textbf{South\_africa} & 65/52 & 57/45 & 81/72 & 79/58 & N/A \\\hline
\textbf{SIRIS} & 39/35 & 35/30 & 50/45 & 47/37 & 43/44 \\\hline
\textbf{Bordingon} & 55/34 & 64/36 & 69/51 & 75/39 & N/A \\\hline
\end{tabular}
\label{table10}
\end{table}

\newpage

\begin{table}[!ht]
\centering
\caption{Recall scores with micro/macro averaging (percentages, \%) for 10 classification methods and 5 validation datasets. Bolded is the best result in the column, asterisk for multiple ``winners'' depending on micro- or macro-averaging.
}
\begin{tabular}{|l|c|c|c|c|c|}
\hline
\textbf{Method \textbackslash~ dataset} & \textbf{Aurora1} & \textbf{Aurora2} & \textbf{Els\_multilabel} & \textbf{Chile} & \textbf{OSDG} \\ \hline
\textbf{Auckland\_v2} & \textbf{63/60} & \textbf{55/48} & \textbf{85/81} & 59/39 & 51/47 \\\hline
\textbf{Aurora\_ml} & 60/59 & 46/48 & 71/69 & \textbf{61/50} & \textbf{65/58} \\\hline
\textbf{Aurora\_v5} & 47/42 & 10/17 & 26/32 & 7/11 & 17/16 \\\hline
\textbf{Els\_2020} & 42/34 & 38/27 & 57/42 & 42/20 & 24/22 \\\hline
\textbf{Els\_2021} & 50/48 & 34/35 & 78/75 & 35/29 & 41/37 \\\hline
\textbf{Els\_2022} & 50/48 & 34/35 & 78/75 & 35/29 & 41/38 \\\hline
\textbf{Els\_2023} & 48/46 & 33/33 & 76/73 & 34/25 & 41/38 \\\hline
\textbf{South\_africa} & 50/44 & 47/42 & 70/60 & 59/42 & N/A \\\hline
\textbf{SIRIS} & 62/60 & 44/47 & 80/78 & 45/43 & 63/58 \\\hline
\textbf{Bordingon} & 50/42 & 53/37 & 64/55 & 61/33 & N/A \\\hline
\end{tabular}
\label{table11}
\end{table}

\begin{table}[!ht]
\centering
\caption{Recall scores with micro/macro averaging (percentages, \%) for 12 classification methods and 5 validation datasets. Here, the validation is performed only against a subset of SDGs for which we have Bergen 2023 queries: 1, 2, 3, 4, 5, 11, 12, 13, 14, and 15.
}
\begin{tabular}{|l|c|c|c|c|c|}
\hline
\textbf{Method \textbackslash~ dataset} & \textbf{Aurora1} & \textbf{Aurora2} & \textbf{Els\_multilabel} & \textbf{Chile} & \textbf{OSDG} \\ \hline
\textbf{Auckland\_v2} & \textbf{66/63} & 71/57* & \textbf{88/83} & 68/46 & 55/56 \\\hline
\textbf{Aurora\_ml} & 58/56 & 47/44 & 71/68 & 65/52 & 62/61 \\\hline
\textbf{Aurora\_v5} & 56/53 & 11/24 & 28/40 & 8/14 & 19/20 \\\hline
\textbf{Bergen\_2023\_baa} & 9/13 & 9/13 & 9/16 & 8/14 & N/A \\\hline
\textbf{Bergen\_2023\_bta} & 12/26 & 12/22 & 14/31 & 10/16 & N/A \\\hline
\textbf{Els\_2020} & 46/36 & 54/34 & 62/44 & 50/26 & 27/27 \\\hline
\textbf{Els\_2021} & 50/47 & 39/40 & 78/75 & 39/33 & 38/38 \\\hline
\textbf{Els\_2022} & 50/47 & 39/40 & 78/75 & 39/33 & 38/39 \\\hline
\textbf{Els\_2023} & 48/45 & 38/38 & 77/72 & 39/29 & 39/40 \\\hline
\textbf{South\_africa} & 56/51 & 59/52 & 78/72 & \textbf{69/57} & N/A \\\hline
\textbf{SIRIS} & 65/62 & 52/53 & 81/80 & 49/48 & \textbf{63/63} \\\hline
\textbf{Bordingon} & 51/40 & 72/46* & 67/55 & 69/42 & N/A \\\hline
\end{tabular}
\label{table12}
\end{table}

\end{document}